\documentclass[preprint]{elsarticle}

\usepackage{array,xspace,multirow,hhline,tikz,colortbl,tabularx,booktabs,fixltx2e,amsmath,amssymb,amsfonts,amsthm}
\usepackage{algorithm}
\usepackage{algorithmic}
\usepackage{verbatim,ifthen}
\usepackage{enumitem}
\usepackage{pifont}
\usepackage{ifthen}
\usepackage{calrsfs,mathrsfs}
\usepackage{bbding,pifont}
\usepackage{pgflibraryshapes}
\usetikzlibrary{trees}
\usetikzlibrary{positioning,chains,fit,shapes,calc}
\usepackage{tkz-graph}

\usepackage{eucal}

\usepackage{tikz}
\usetikzlibrary{arrows}
\usetikzlibrary{decorations.pathreplacing}

	\usepackage{varioref}

\definecolor{light-gray}{gray}{0.9}

	\newtheorem{proposition}{Proposition}%
	\newtheorem{example}{Example}
		
	\newcommand\eat[1]{}

	\usepackage{enumitem}
	\setenumerate[1]{label=\rm(\it{\roman{*}}\rm),ref=({\it\roman{*}}),leftmargin=*}
	\newlength{\wordlength}
	\newcommand{\wordbox}[3][c]{\settowidth{\wordlength}{#3}\makebox[\wordlength][#1]{#2}}

	\newcommand{\eqclass}[2][]{\ifthenelse{\equal{#1}{}}{[#2]}{[#2]_{\sim_{#1}}}}

	\newcommand{\pref}{\succsim \xspace}
	\newcommand{\Pref}[1][]{
		\ifthenelse{\equal{#1}{}}{\mathrel R}{\mathop{R_{#1}}}
	}                                          
	\newcommand{\sPref}[1][]{                  
		\ifthenelse{\equal{#1}{}}{\mathrel P}{\mathop{P_{#1}}}
	}                                          
	\newcommand{\Indiff}[1][]{                 
		\ifthenelse{\equal{#1}{}}{\mathrel I}{\mathop{I_{#1}}}
	}
	\newcommand{\prefset}[1][]{\ifthenelse{\equal{#1}{}}{\mathcal{R}}{\mathcal{R}_{#1}}}

\usepackage{enumitem}
\setenumerate[1]{label=\rm(\it{\roman{*}}\rm),ref=({\it\roman{*}}),leftmargin=*}

\newcommand{\nbh}[1][]{
	\ifthenelse{\equal{#1}{}}{\nu}{\nu(#1)}
}

\newcommand{\cstr}[1][]{
	\ifthenelse{\equal{#1}{}}{\mathscr S}{\cstr(#1)}
}

\newcommand{\choice}[1][]{
	\ifthenelse{\equal{#1}{}}{\mathit{C}}{\choice(#1)}
}

\usepackage{boxedminipage}
\usepackage{xspace}

\tikzset{
  treenode/.style = {align=center, inner sep=0pt, text centered,
    font=\sffamily},
  arn_n/.style = {treenode, circle, white, font=\sffamily\bfseries, draw=black,
    fill=black, text width=1.5em},
  arn_r/.style = {treenode, circle, red, draw=red, 
    text width=1.5em, very thick},
  arn_x/.style = {treenode, rectangle, draw=black,
    minimum width=0.5em, minimum height=0.5em}
}

\begin{document}

\title{A note on the undercut procedure}

\author{Haris Aziz}
	\ead{haris.aziz@nicta.com.au}
	\address{NICTA and UNSW, Kensington 2033, Australia}



\begin{abstract}
	The undercut procedure was presented by \citet{BKK12a} as a procedure for identifying an envy-free allocation when agents have preferences over sets of objects. They assumed that agents have strict preferences over objects and their preferences are extended over to sets of objects via the responsive set extension. We point out some shortcomings of the undercut procedure. We then simplify the undercut procedure of \citet{BKK12a} and show that it works under a more general condition where agents may express indifference between objects and they may not necessarily have responsive preferences over sets of objects. Finally, we show that the procedure works even if agents have unequal claims.
\end{abstract}


\maketitle

\section{Introduction}

Allocation of indivisible resources is one of the most fundamental problems in \emph{fair division} and \emph{multiagent resource allocation}~\citep{BrTa96a}. 
Many of the fair division settings feature two agents since  disputes often concern two parties. Recently \citet{BKK12a} presented the \emph{undercut procedure} which is an elegant procedure to divide a set of contested indivisible objects fairly among two agents. A crucial assumption in the paper was that agents have a strict ranking over the objects and the preferences over \emph{sets} of objects are \emph{responsive}. Preferences over sets of objects are responsive if for any set in which an object is replaced by a more preferred object, the new set is more preferred. We first show that the assumption of responsive preferences can be somewhat restrictive. We also identify three shortcomings of the first few steps of the undercut procedure. Finally we rectify the shortcomings by simplifying the undercut procedure of  \citet{BKK12a} and showing that it returns an envy-free allocation (if it exists) under a more general preference restriction called \emph{separability}.

The setting we consider concerns two agents $1$ and $2$ and a set of objects $O$. Both agents have \emph{complete} and \emph{transitive} preferences $\pref_1$ and $\pref_2$ over the subsets of objects in $O$. 
The goal is to identify an envy-free split $(S,-S)$ where $S$ is the allocation of agent $1$ and $-S=O\setminus S$ is the allocation of agent $2$.

\section{The undercut procedure}

The undercut procedure is a discrete generalization of the divide and choose cake cutting protocol~\citep[Chapter 1, ][]{BrTa96a}. 
The elegance of the undercut procedure lies in the fact that although agents have preferences over sets of objects, it is sufficient  to only consider or query about the \emph{minimal bundles} of the agents. 
A subset $S\subseteq O$ is a \emph{minimal bundle} for agent $i$ if $S\pref_i -S$ and for any $T\subset S$, $-T\succ_i T$.
The set of minimal bundles of agent $i$ is denoted by $MB_i$. 
Any envy-free split $(S,-S)$ of $O$ is \emph{trivial} if $S\sim_1 -S$ and $-S\sim_2 S$.
The main idea underlying the undercut procedure is that there exists a non-trivial envy-free allocation if the set of minimal bundles of both agents is not the same. 
The undercut procedure goes through the minimal bundles of the agents to identify an envy-free allocation if it exists~\citep{BKK12a}.  We refer the reader to Algorithm~\ref{algo:undercut} for an adapted specification of the undercut procedure. 

\begin{algorithm}[h!]
	  \caption{Undercut procedure of \citet{BKK12a}}
	  \label{algo:undercut}
	\footnotesize
	\renewcommand{\algorithmicrequire}{\wordbox[l]{\textbf{Input}:}{\textbf{Output}:}} 
	 \renewcommand{\algorithmicensure}{\wordbox[l]{\textbf{Output}:}{\textbf{Output}:}}
	\begin{algorithmic}
		\REQUIRE $(N,O,\pref)$
		\ENSURE Envy-free split if it is exists.
	\end{algorithmic}
	\algsetup{linenodelimiter=\,}
	  \begin{algorithmic}[1] 
	\small
			\STATE \textbf{Generation Phase:} Agent $1$ and $2$'s most preferred objects are given to them if they do not coincide. If the object coincides, then it is placed in the contested pile $I_C\subset O$. The process continues until all objects have been names by at least one agent. If the contested pile is empty, the procedure ends. Otherwise, each agent $i$ identifies his  set of minimal bundles $MB_i$ of $I_C$.
			\STATE If $MB_1 \neq MB_2$, each agent $i$ provides reports to the mechanism a ranking of his  minimal bundles. An agent $i$ is chosen at random, and one of $i$'s top-ranked minimal bundle $S$ is considered. If  $S\notin MB_{-i}$ , then it becomes the proposal, and $i$ is the proposer. If  $S\in MB_{-i}$, then one of $-i$'s top-ranked minimal bundle $S'$ is considered. If $S'\notin MB_i$, then it becomes the proposal, and $-i$ the proposer. If  $S'\in MB_i$, then the process continues until a minimal bundle of one agent is found that is not a minimal bundle of the other. Then proceed to step 4.
			\STATE If $MB_1 = MB_2$, and there exists an $S$ such that $S \in  MB_i$ and $−S \in MB_i$ (and, therefore $S,-S \in MB_{-i}$ also), then $S$ becomes the proposal. If there is no minimal bundle $S$ such that $−S$ is also a minimal bundle, then a minimal bundle is chosen randomly and becomes the proposal.
				\STATE Assume that $S$ is the proposal and the proposer is $i$. Then $-i$ may respond by
	$i)$ accepting $-S$ of $I_C$
or $ii)$ undercutting $i$'s proposal, i.e., taking his most-preferred subset $T$ and giving $-T$ to $-i$.
				%
				%
				The procedure ends. An agent's subset of $O$ consists of all objects received in steps 1 and 2, plus the agent's share of the contested pile determined in step 4.
			

	 \end{algorithmic}
	\label{algo:subroutine}
	\end{algorithm}

\paragraph{Limitation of responsive preferences}

The undercut procedure was shown to find an envy-free allocation if the preferences of agents are \emph{responsive}. 
\emph{Responsiveness} is a  well-established preference restriction on preferences over sets of objects which assumes that the agents have preferences over the individual objects. 
Preferences over sets of objects are \emph{responsive}, if for any two sets that differ only in one object, the agent prefers the set containing the more preferred object~\citep{BBP04a}.
We first highlight that responsive preferences can be restrictive. 
\begin{example}
In a divorce dispute, husband $h$ may prefer each of the two family dogs $d_1$ and $d_2$ over the car $c$: 
$\{d_1\} \succ_h \{d_2\} \succ_h \{c\} \succ_h \emptyset.$ If the husband's preferences are responsive, then his preferences over the set of issues is as follows:
$\{d_1,d_2,c\}  \succ_h \{d_1,d_2\}  \succ_h \{d_1,c\} \succ_h  \{d_2,c\}. 
$

However it may be the case that the husband prefers the set of a car and a dog to the set of two dogs: $\{d_1,c\}\succ_h \{d_1,d_2\}$. This way he will have both a companion and a ride. $\{d_1,d_2,c\}  \succ_h \{d_1,c\}  \succ_h \{d_2,c\} \succ_h  \{d_1,d_2\}.$
\end{example}

A preference relation $\succ$ is \emph{separable} if for all
$S\subset O$ such that $x\notin S$, the following holds: 
 $\{x\}\succ \emptyset$ if and only if $S\cup \{x\} \succ S$~\citep{BBP04a}. 
Informally, separability means that if an agent prefers having the object than having nothing, he would also prefer the inclusion of the object in any other set that does not include the object. 
Whereas responsive preferences are separable, separable preferences are more general than responsive preferences.  Just as in \citep{BKK12a}, we will assume that all the objects are desirable. However, we will not use the restriction in  \citep{BKK12a} that preferences over objects do not admit ties.

\paragraph{Issues with the generation phase of the undercut procedure}

In the generation phase of the undercut procedure (Algorithm~\ref{algo:undercut}), each agent sequentially picks up his maximal object if it is uncontested. Otherwise each contested object goes into the \emph{``contested pile''}.
We argue that the generation phase of the undercut procedure (also referred to as the \emph{generation phase} in \citep{VeKi13a}) has some drawbacks.
\emph{Firstly},	undercut may fail to identify an envy-free split because of the generation phase.
Let us consider the following preferences of agents 1 and 2: $a \mathrel{\succ_1} b \mathrel{\succ_1} c \mathrel{\succ_1} d$ and $b \mathrel{\succ_2} c \mathrel{\succ_2} d \mathrel{\succ_2} a.$
		If  $\{a,d\} \sim_1 \{b,c\}$, we know that the assignment which allocated $\{a,d\}$ to agent $1$ and $\{b,c\}$ to agent $2$ is envy-free. However undercut fails to compute this assignment. The reason is that in the generation phase, agent $1$ takes $a$ and agent $2$ takes $b$. After this the contested pile is $\{c,d\}$. The undercut procedure ends up in a deadlock in this contested pile. 
\emph{Secondly}, even if the undercut procedure works for certain responsive preferences, the generation phase hinders it from working for separable preferences.
\emph{Thirdly}, the generation phase use sequential allocation. It is well-understood that sequential allocation is highly susceptible to manipulation if at least one agent has sufficient information about the other agent's preferences~\citep{KoCh71a,VeKi13a}. 
One point which goes in favour of Step 1 and 2 of the undercut procedure is that it decreases the size of the contested pile which was presumably the motivation behind the steps. 
\paragraph{Simplified undercut procedure}

Next, we show that the simplified undercut procedure works for transitive and separable preferences. 
We define a \textit{simplified undercut} procedure as follows.
\textbf{\emph{Simplified Undercut:}}	\emph{Treat the set of all objects as the contested pile and run the original undercut procedure while ignoring the generation phase of the original undercut procedure.}

\begin{proposition}
		For transitive and separable preferences, there is a non-trivial envy-free split if and only if the set of minimal bundles of both agents is not the same. Furthermore the simplified undercut procedure finds such an a split.  
\end{proposition}
\begin{proof}
	The argument is similar to the one for the proof of \citep[Theorem 1, ][]{BKK12a}. 
	We first prove that if a  non-trivial envy-free split exists it implies that the set of minimal bundles of both agents is not the same. Let us assume that a non-trivial envy-free split $(S,-S)$ exists. 
Then there must be an agent $i\in \{1,2\}$ such that $S\succ_i -S$. By the definition of minimal bundle, we know that $-S\notin {MB}_i$. Without loss of generality, we can assume that $S\in {MB}_{i}$. If it were not then we argue that there exists an $S'$ such that $S'\in {MB}_{i}$, $S'\subset S$, such that $S'\succsim_{i} -S'$. 
For an $S'\subset S$, by separability, we know that $S\succ_i S'$. Similarly, by separability, we know that $-S'\succ_i -S$ because $-S'$ can be obtained from $-S$ by adding those elements to $-S$ as the elements that are removed from $S$ to obtain $S'$. Since $\pref_i$ is transitive and complete, there exists some $S'$ such that $S'\succsim_i -S'$ for which there exists no subset $S''\subset S'$ such that $S''\pref_i -S''$.

Now if $S'\sim -S'$, then we know that $S'\in {MB}_{i}$. By separability, we also know that $-S'\succ_{-i} -S \succsim_{-i} S \succ_{-i} S'$. Hence $S'\notin {MB}_{-i}$ which means that  set of minimal bundles of both agents is not the same.
If $S\notin {MB}_{-i}$, then we have already proved that the set of minimal bundles of the two agents are different. Now let us assume that $S\in {MB}_{-i}$. Then we know that $S\succsim_{-i} -S$. Since $(S,-S)$ is envy-free, then it follows that $-S\succsim_{-i} S$. Hence $S\sim_{-i}-S$. If $-S\in {MB}_{-i}$ then we are already done. We show that $-S$ is indeed in ${MB}_{-i}$. Consider any $T\subset-S$ which implies by separability that $-S\succ_{-i} T$.
Since $S=O \setminus -S$ and since $T\subset-S$, we know that $O\setminus T = -T\supset S$. This implies by separability that $-T\succ_{-i} S$. Since $-T\succ_{-i} S$, $-S\succ_{-i} T$ and $S\sim_{-i}-S$, we get by transitivity that $-T\succ_{-i} T$. 
Hence we have shown that $-S\in {MB}_{-i}$. Since we know that $-S\notin {MB}_{i}$, the set of minimal bundles of both agents is not the same.

	We now prove that if the two agents do not have the same set of minimal bundles then there exists a non-trivial envy-free split.
	Let us assume that the two agents do not have the same set of minimal bundles i.e., there exists an $S\subset O$ such that $S\in {MB}_{i}$ and $S\notin {MB}_{-i}$. If $-S\succsim_{-i} S$, then $(S,-S)$ is an envy-free split. 	When agent $i$ will propose $(S,-S)$, agent $-i$ will accept it. Therefore let us look at the other case when $S\succ_{-i} -S.$
If $S\succ_{-i} -S$,  and  $S\notin {MB}_{-i}$, then by the definition of a minimal bundle we know that there exists a $T\subset S$ such that $T\succsim_{-i}-T$. Agent $-i$ will undercut the proposal $(S,-S)$ of agent $i$ and will be ready to take $T$. 
For agent $1$, we know that $S\succ_i -T \succ_i T\succ_i -S.$ Thus $(-T,T)$ is an envy-free split.
\end{proof}


The argument is similar to the one for the proof of \citep[Theorem 1, ][]{BKK12a}. 
If a trivial envy-free split exists, the simplified undercut procedure will find it since it considers the minimal bundles of the two agents. If a trivial envy-free split does not exist but a non-trivial one does, even then the simplified undercut procedure will find it.
If agents have unequal claims say claim $c_i$ for agent $i$,  then the definition of envy-freeness can be easily extended as follows: $u_i(S)\geq \frac{c_i}{c_{-i}}u_i(-S)$ for an allocation where $i$ gets $S$.
If agents have unequal claims, the undercut procedure still works as follows. We simply redefine a minimal bundle $S$ for agent $i$ as a set of objects such that $u_i(S)\geq \frac{c_i}{c_{-i}}u_{i}(-S)$ and for any $T\subset S$, $u_i(T)< \frac{c_i}{c_{-i}}u_{i}(-T)$. 

\section*{Acknowledgment}
NICTA is funded by the Australian Government through the Department of Communications and the Australian Research Council through the ICT Centre of Excellence Program.


 \end{document}